\documentclass{article}
\usepackage{graphicx,caption}
\usepackage{fullpage}
\usepackage{fancyhdr,amsmath,amsthm,amssymb,bm,url,enumerate,float}
\usepackage{xcolor}
\usepackage{appendix}
\usepackage{comment}
\usepackage{authblk}
\usepackage{setspace}
\usepackage{tikz}
\usepackage{subcaption}
\usetikzlibrary{arrows.meta,shapes}

\DeclareMathOperator*{\logit}{logit}
\newcommand{\E}{\mathbb{E}}

\newcommand{\keywords}[1]{\textbf{\textit{Keywords---}} #1}

\title{Time-smoothed inverse probability weighted estimation of effects of generalized time-varying treatment strategies on repeated outcomes truncated by death}

\author[1,2]{Sean McGrath}
\author[3]{Takuya Kawahara}
\author[2]{Joshua Petimar}
\author[2]{Sheryl L.\ Rifas-Shiman}
\author[4]{Iván Díaz}
\author[2]{Jason P.\ Block}
\author[2]{Jessica G.\ Young}

\affil[1]{\small Department of Biostatistics, Yale University, New Haven, CT, USA}
\affil[2]{\small Department of Population Medicine, Harvard Medical School and Harvard Pilgrim Health Care Institute, Boston, MA, USA}
\affil[3]{\small Clinical Research Promotion Center, The University of Tokyo Hospital, Japan}
\affil[4]{\small Division of Biostatistics, New York University, New York, NY, USA}
\date{}

\renewcommand{\addcontentsline}[3]{}

\begin{document}

\maketitle

\begin{abstract}
Researchers are often interested in estimating effects of generalized time-varying treatment strategies on the mean of an outcome at one or more selected follow-up times of interest. For example, the Medications and Weight Gain in PCORnet (MedWeight) study aimed to estimate effects of adhering to flexible medication regimes on future weight change using electronic health records (EHR) data. This problem presents several methodological challenges that have not been jointly addressed in the prior literature. First, this setting involves treatment strategies that vary over time and depend dynamically and non-deterministically on measured confounder history. Second, the outcome is repeatedly, non-monotonically, informatively, and sparsely measured in the data source. Third, some individuals die during follow-up, rendering the outcome of interest undefined at the follow-up time of interest. In this article, we pose a range of inverse probability weighted (IPW) estimators targeting effects of generalized time-varying treatment strategies in truncation by death settings that allow time-smoothing for precision gain. We conducted simulation studies that confirm precision gains of the time-smoothed IPW approaches over more conventional IPW approaches that do not leverage the repeated outcome measurements. We illustrate an application of the IPW approaches to estimate comparative effects of adhering to flexible antidepressant medication strategies on future weight change. The methods are implemented in the accompanying R package, \verb|smoothedIPW|.

\end{abstract}

\keywords{causal inference, missing data, generalized time-varying treatment strategies, truncation by death, inverse probability weighting, marginal structural models, repeated outcomes}

\section{Introduction}

Researchers are often interested in using longitudinal data to estimate causal effects of time-varying treatment strategies on the mean of an outcome at one or more selected follow-up times of interest.  For example, the Medications and Weight Gain in PCORnet (MedWeight) Study aimed to leverage electronic health record (EHR) data to inform comparative effects of initiating, and subsequently adhering to, different medications within the same class (e.g., antidepressants) on mean weight change 6, 12, and 24 months post-baseline.  There are several complex features of the MedWeight study attributable to a combination of the nature of this type of causal question and the nature of this type of data source.  

First, the treatment strategies of interest are not only time-varying, such that measured time-varying confounders may be affected by treatment \cite{hernan2020what}, but they are also defined such that treatment assignment under the strategy at a given time depends \textsl{dynamically} and \textsl{non-deterministically} on these measured confounders; for example, allowing grace periods for periodic treatment breaks and stopping or continuing medication as per doctor and/or patient discretion upon an event like pregnancy. Second, the outcome of interest in this study (weight change) is one of many examples of a variable that is \textsl{repeatedly, non-monotonically, informatively, and sparsely} measured in an EHR: \textsl{repeatedly} in that an individual meeting eligibility criteria for the analysis at baseline may have multiple subsequent measures of weight change available in the longitudinal data; \textsl{non-monotonically} in that an individual may have a measure of weight change at 5 months after baseline but not at 4 months or at 6 months;  \textsl{informatively} in that there may exist (measured and unmeasured) shared causes of whether an individual has a measure of weight change and its actual value; and \textsl{sparsely} in that there may be few individuals with weight change measured at any individually selected follow-up time (e.g., 6 months) relative to the total eligible sample size at baseline.  Sparse measurement, in particular, makes analytic approaches that can smooth over all available measures of weight change attractive for precision gain.  Third, weight change is an example of an outcome that is truncated by death in that this outcome is undefined (i.e., does not have a substantively meaningful value) at any time after an individual has died.  

There is a vast literature that has considered, at least, the first two complexities separately, but not jointly.  Specifically, it is now well-established that a counterfactual outcome mean indexed by either a static deterministic strategy (e.g., ``always treat'') or a pragmatic dynamic, possibly stochastic, treatment rule dependent on past measured covariates in the study can be identified under conditions that are expected to hold in a sequentially randomized trial, but also more broadly, via Robins's \textsl{generalized g-formula} \cite{robins1986new}. It has been further shown that the class of strategies identified by the generalized g-formula is even broader, including strategies that may further depend on the history of the patient's so-called \textsl{natural treatment value} \cite{richardson2013single,young2014identification}. Different estimators have been developed for the generalized g-formula indexed by any time-varying treatment strategy in this broad class and contrasts comparing different strategies to target causal effects. These include the parametric g-formula \cite{robins1986new, mcgrath2020gformula, wen2021parametric, taubman2009intervening, WHOchap}, inverse probability weighting \cite{young2018inverse}, and various double debiased methods \cite{lendle2017ltmle, diaz2023nonparametric, wen2023intervention}. In the literature on double debiased estimation, this class of strategies identified by the generalized g-formula has been termed \textsl{longitudinal modified treatment policies} (LMTPs). However, regardless of estimator, to date, the literature on estimation of a generalized g-formula in order to target the mean of an outcome, like weight change, under a selected strategy at a selected follow-up time limits utilization of outcome data at that selected time and does not incorporate any time smoothing to leverage repeatedly available outcome measurements. 

Conversely, there is an even larger literature on statistical approaches for estimating parameters of models for repeatedly measured outcomes \cite{hardin2002generalized, stroup2024generalized} with further extensions for causal inference methods that account for non-monotonic and informative measurement \cite{robins1995analysis} as well as time-varying treatment strategies.  However, this is restricted to stricter subclasses of those accommodated by the generalized g-formula: static deterministic strategies \cite{cole2008constructing, cole2007determining} and dynamic deterministic strategies \cite{hu2019causal}. To our knowledge, there does not currently exist more general methodology for targeting any causal effect within this broader class of strategies, including stochastic dynamic strategies and strategies dependent on natural treatment values, that simultaneously accommodates time-smoothing for this type of outcome.  

The challenge of truncation by death is more fundamental than the other two outlined above. When some individuals in the population die prior to a selected outcome time of interest $t^*$ and the outcome is undefined at that time for an individual who has died between baseline and $t^*$, the total population average causal effect of the desired treatment strategies on the outcome at $t^*$ is itself undefined.  Therefore, in this setting, we do not have the option to ground causal inferences in this default notion of a total effect -- the only notion that is even in principle identified under assumptions guaranteed by design in an ideally executed randomized trial.  

Solutions to this problem have included assigning an extreme value to the outcome for dead individuals and considering causal effects on a quantile such as the median (as opposed to a mean) \cite{hu2019causal, xiang2023survival, xiang2024estimating}. While this maintains the identification advantages of a total effect, it may not correctly translate the investigator's (often implicit) ``causal story'' \cite{young2024story} clarifying what they actually wish to learn from the analysis. In the literature explicitly considering identification via the generalized g-formula for an outcome subject to truncation by death, the most common ``solution'' has been to utilize methods justified for comparative effects of the specified treatment strategies under an additional, unspecified intervention that would somehow fully ``eliminate death'', coinciding with a special case of a controlled direct effect\cite{robins1992identifiability}.  For such an estimand, death can be understood as a right-censoring process \cite{young2020causal}. In addition to requiring additional assumptions on this censoring process that include ``no unmeasured selection factors'' for death and the outcome to achieve identification,  this causal target will rarely, if ever, correctly translate what is motivating a clinical investigator. 

Many authors have advocated grounding causal inferences in the survivor average causal effect (SACE) \cite{robins1986new, frangakis2002principal} for truncation by death settings for which existing results are largely limited to time-fixed treatments and do not accommodate generalized time-varying treatment strategies. The SACE quantifies the causal treatment effect in the subpopulation of individuals who would survive to the outcome time of interest regardless of treatment.  Like the controlled direct effect, the SACE will require assumptions for identification beyond the total effect.  It also references a subpopulation that is never directly observable by its ``cross-world'' nature: we can never directly observe who would survive regardless of treatment because we can only observe an individual's survival status under the treatment strategy they actually followed. 

Stensrud et al. \cite{stensrud2023conditional} recently posed an alternative notion of causal effect for truncation by death settings: a conditional separable effect.  Grounding causal inference in a conditional separable effect requires assuming modifications to the current measured study treatments such that, informally, the modified treatments would have ``no effect on death''.  A conditional separable effect is then defined as what would be consistently estimated by a comparison of outcomes across treatment arms among only survivors in a trial in which the modified treatments are randomized.  The assumption of ``no effect on death'' endows this contrast with a property that any notion of causal effect should arguably have: comparing outcomes across different treatment strategies but in ``comparable'' populations.  A subtlety is that, depending on the causal ontology we are working under \cite{robins2010alternative, sarvet2022without}, the assumption of ``no treatment effect on death'' could simply refer to no distributional difference or a sharp null of no individual-level effect.  Under the latter, a conditional separable effect will in fact coincide with a SACE for the modified treatment comparison. 

In this paper, we pose a range of inverse probability weighted estimators of the generalized g-formula to target effects of generalized treatment strategies/LMTPs that allow time-smoothing for precision gain. We provide estimators that can target a controlled direct effect under ``elimination of death'' along with an alternative effect that constitutes a trivial case of a conditional separable effect -- the case where the study treatment requires no modification to make the ``no effect on death'' assumption hold.  We argue that this is a reasonable assumption for choices of $t^*$ closer to baseline in the MedWeight study and, further, that this latter effect notion best translates the underlying causal story of MedWeight precisely for these earlier choices of outcome time. Our presentation more broadly reinforces the importance of the outcome time selection in ensuring the correct interpretation of a counterfactual target and its identification, particularly in truncation by death settings where leveraging all repeated sparsely measured outcomes may be desired for increasing statistical precision but may not all be of substantive/"causal" interest.

\section{Observed data structure}\label{sec: obsdata}
Consider a longitudinal study containing measurements on a sample of $n$ individuals from a study population of interest (e.g. treatment naive individuals with a new depression diagnosis and no treatment contraindications making a choice between initiating one of two common antidepressant monotherapies with their clinician, sertraline or citalopram). Let $t = 0,1,2,\ldots,\tau$ denote discrete follow-up intervals of equal length (e.g. months) containing time-updated measurements where $t=0$ indicates the baseline interval (e.g. the month of this antidepressant monotherapy initiation decision) and $t = \tau$ denotes the maximum interval for which measurements are available in the data set (e.g. 24 months). 

Let $A_t$ denote a measured treatment in interval $t$, meaning a variable that the investigator would ideally intervene on to answer their causal question; e.g. $A_0$ an indicator of which antidepressant was actually initiated ($A_0=1$ if citalopram and $A_0=0$ if sertraline) and $A_t$ an indicator of whether the patient was still taking the treatment they initiated at $t>0$. Throughout we will use the notational conventions of an overbar to denote the history of a random variable between baseline and a specified time $t$, e.g., $\bar{A}_t := (A_0, A_1, \dots, A_t)$, and an underbar to denote the future of a random variable between a specified time $t$ and $t=\tau$, e.g., $\underline{A}_t := (A_t, A_1, \dots, A_\tau)$.  For $t>0$, let $Y_{t}$ denote a factual, but not necessarily measured/observed, outcome in interval $t$ (e.g. weight change in kg at $t$ relative to weight at baseline, $t=0$) and $R_t$ an indicator of whether $Y_t$ was measured \textsl{at} time $t$. When $R_t=0$, the value of $Y_t$ is unknown to the investigator/analyst.  $R_t$ quantifies a non-monotonic event process meaning that there is no determinism between the value of $R_t$ and the value of $R_s$, $s\neq t$ (e.g. an individual can have a measure of weight change at times $t=2,6,13$ and no measure of weight change at any other time in the data set). We let $L_t$ denote a vector of other measured covariates in interval $t$ (e.g. for $t=0$, baseline weight in kg, sex, race/ethnicity; for $t>0$, any new diagnosis by $t$ that contraindicates continuing treatment at $t$, most recent lab measures relative to $t$, recent engagement with the health system at $t$).  $\overline{L}_t$ implicitly includes $\overline{R}_{t-1}$. 

Let $C_t$, $t>0$ be an indicator of ``loss to follow-up'' \textsl{by} time $t$. We will define an individual as ``lost to follow-up'' by a time $t$ if their data is classified as ``unreliable'' from that time forward. For example, in studies utilizing EHR data from a health system, an individual might be classified as lost to follow-up the first time that they have gone a specified number of months with no evidence of engagement with that health system.  Unlike $R_t$, the variable $C_t$ quantifies a monotonic process, meaning that if $C_t=1$ then $\underline{C}_{t+1}=1$ and if $C_t=0$ then $\overline{C}_{t-1}=0$. Finally, we define $D_t$ as an indicator of death \textsl{by} time $t$, which is also a monotonic event process.  Importantly, when $D_t=1$, the \textsl{factual} values of all $\underline{Y}_t$ are undefined (i.e., considered meaningless by the investigator).   Until Section \ref{sec: identification deaths}, we will make the simplifying assumption that no deaths can occur in this study population between baseline and $t=\tau$ ($\overline{D}_\tau \equiv 0$). Note that, our subsequent formalization of how the measured study variables under this observed data structure can be used to estimate certain notions of causal effect rests \textsl{either} on the assumption of a temporal ordering within each interval $t=0,\ldots,\tau$ of $(L_t,A_t,D_t,C_t,R_t,Y_t)$ \textsl{or} that time interval lengths are selected small enough that there is no dependence between events within an interval $t$.

\section{A definition of causal effect when deaths do not occur}\label{sec: question}

Let $t^*$ denote a selected outcome time \textsl{of interest to the investigator}, $1\leq t^*\leq \tau$. Further, let $X_t$ denote an observed indicator of what we will term a \textsl{censoring event} by $t=1,\ldots,t^*$ with $X_0\equiv 0$.  We define the components of $X_t$ to include $C_t$ for all $t=1,\ldots,t^*$, We further include $(1-R_t)$ in $X_t$ \textsl{only} for $t=t^*$. In turn, define $Y_{t^*}^{g}$ as an individual's outcome value at that follow-up time of interest $t=t^*$ had, possibly contrary to fact, they adhered to a time-varying treatment strategy $g$ \textsl{of interest to the investigator} up to $t^*$ and also remained uncensored through $t^*$. Following Richardson and Robins \cite{richardson2013single}, we further define $L^{g}_t$, $A_t^{g}$, $X^{g}_t$ as the \textsl{natural counterfactual values} of the measured study covariates, treatment, and censoring at $t\leq t^*$ indexed by $g$ defined as the values of these variables at $t$ under adherence to $g$ and censoring elimination up to, but not including, time $t$.  In contrast, to the natural treatment value $A_t^g$, define $A^{g+}_t$ and $X^{g+}_t$ as the \textsl{intervention values} of treatment and censoring under adherence to $g$ through time $t$ and censoring elimination.  By definition, $\overline{X}^{g+}_{t^*}=0$.  By contrast, the value of $A^{g+}_t$ at any $t=0,\ldots,t^*$ will depend on the choice of $g$.  Our presentation restricts consideration to settings where interest lies only in time-varying treatments strategies $g$ defined such that the intervention treatment value at $t$, $A^{g+}_t$ at a time $t$ between baseline and $t^*$ can, at most, depend on $(\overline{A}^{g+}_{t-1},\overline{L}^g_t, \overline{A}^g_t$), i.e., strategies where treatment assignment can at most depend on the past natural values of measured covariates and past natural and intervention values of the treatment itself.  These strategies can depend deterministically on $(\overline{A}^{g+}_{t-1},\overline{L}^g_t, \overline{A}^g_t$) \textsl{or} \textsl{stochastically} (coinciding with a deterministic strategy conditional on an exogenous randomizer \cite{wanis2024grace}).  Strategies within this class have also been referred to as longitudinal modified treatment policies (LMTPs) \cite{diaz2023nonparametric}. We will denote the set of all strategies in this class $\mathcal{G}$.  

In turn, we can define the mean difference
\begin{equation}
\mbox{E}[Y_{t^*}^{g_1}]-\mbox{E}[Y_{t^*}^{g_0}]\label{total}
\end{equation}
indexed any two strategies $g_1, g_0 \in \mathcal{G}$, $g_1\neq g_0$ and refer to this mean difference as the population average \textsl{causal effect} of the treatment strategy $g_1$ versus $g_0$ on the outcome at time $t^*$ under censoring elimination (i.e., no loss to follow-up through $t^*$ and complete outcome measurement at $t^*$).\footnote{Note that our classification of $(1-R_t^*)$ as ``censoring'' differs slightly from traditional useage of the term.  Here, this classification is due to the fact that when $(1-R_t{^*})=1$ we will not observe the outcome of interest $Y_{t^*}^{g}$, specific to the time $t=t^*$.  However, $(1-R_t{^*})=1$ here does \textsl{not} imply that no more data will be used for that individual after that time in the analysis.  As we will see in Section \ref{sec: time smoothed estimation}, this data will still be used in the context of our time-smoothed estimator.  By expanding $X_t$ to include $(1-R_t)$ at all times $t=1,\ldots,t^*$, rather than only $t=t^*$, this indicator will meet a more traditional understanding of a censoring event where no more data is used after that event occurs, even with our time-smoothed estimator.  In turn this may alter the specific interpretation of \eqref{total} and the nature of identifying assumptions outlined in the next section with regard to censoring. It can also result in substantial precision loss.}  We can alternatively refer to \eqref{total} as a \textsl{total} effect of $g_1$ versus $g_0$ on the population mean outcome at $t^*$ in that \eqref{total} does not isolate any particular mechanism by which $g_1$ versus $g_0$ affects this selected outcome \cite{robins1992identifiability,young2020causal,young2024story}.

Returning to our antidepressant example, consider the following concrete examples of $g_1$ and $g_0$ of interest in the MedWeight study which constitute examples of \textsl{natural grace period strategies} indexed by a choice of grace period length $m$ \cite{wanis2024grace}.  First, for $t=1,\ldots,t^*$, let $\tilde{L}^g_t$ denote a component of $L^g_t$ indicating whether the patient has become pregnant or received bariatric surgery by interval $t$ (both possible treatment contraindications strongly associated with weight change) under the strategy $g$.  Also, for $t=1,\ldots,t^*$, define $G^{g,m}_t$ as an indicator of whether an individual: (i) has reached the end of a \textsl{treatment grace period} under $g$ at time $t$, defined as going $m$ consecutive time intervals without taking their initiated medication under $g$ $(\sum_{s=t-m}^{t-1}[1-A^{g+}_s]=m)$ \textsl{and} (ii) has not yet experienced the possibly contraindicating events of pregnancy or bariatric surgery by $t$ ($\tilde{L}^g_t=0$).  In turn, we define $g_1$ and $g_0$ as follows:
\begin{itemize}
    \item ($g_1$) Initiate citalopram at $t=0$ ($A^{g+}_0=1$). Then at each subsequent $t=1,\ldots,t^*$: if $G^{g,m}_t=0$, the patient can continue to take citalopram or not based on their own decision in consultation with their health care provider, i.e. receive their natural treatment value at $t$ ($A^{g+}_t=A^{g}_t$); otherwise, if $G^{g,m}_t=1$, they must take citalopram in interval $t$ ($A^{g+}_t=1$)
    \item ($g_0$)  Initiate sertraline at $t=0$ ($A^{g+}_0=0$). Then at each subsequent $t=1,\ldots,t^*$: if $G^{g,m}_t=0$, the patient can continue to take sertraline or not based on their own decision in consultation with their health care provider, i.e. receive their natural treatment value at $t$ ($A^{g+}_t=A^{g}_t$); otherwise, if $G^{g,m}_t=1$, they must take sertraline in interval $t$ ($A^{g+}_t=1$)
\end{itemize}
See \cite{cain2010start, hiv2011initiate, lodi2015comparative} for other examples of strategies $g$ that fall within the class defined above.
 
We note here three important features of the contrast \eqref{total}.  First, we can allow two subtly different interpretations of this contrast depending on our causal ontology \cite{robins2010alternative,sarvet2022without}.  Under a \textsl{counterfactual} causal model, we will understand this contrast as the mean of the outcome at the selected time $t^*$ had all individuals in the study population adhered to strategy $g_1$ versus, instead, had all of these same individuals adhered to $g_0$ through $t^*$, again, in both scenarios under censoring elimination.  Under this interpretation, the difference \eqref{total} arguably has the desirable property that it compares outcome means \textsl{under different treatment strategies} but in the \textsl{same population}.  By this property, any difference can \textsl{only} be explained by differences in the strategies and not by comparing different populations (sets of people).  Alternatively, we could take an \textsl{agnostic} position on the existence of individual-level counterfactual outcomes, and understand this contrast as simply the population quantity that the difference in sample means across treatment arms would converge to in a trial where individuals were randomized at baseline to treatment arms where the protocol requires adherence to either $g_1$ or $g_0$ and no censoring, respectively, and everyone perfectly adheres.  Under this agnostic interpretation, the difference \eqref{total} still has the desirable property that any difference cannot be explained by comparing ``non-comparable'' populations/sets of people, with ``non-comparable'' formally defined using d-separation rules relative to a causal diagram representing the agnostic causal model.  We will consider this distinction further in Section \ref{sec: identification deaths}.  

Second, the contrast \eqref{total} is guaranteed identified in an ``ideal trial'' where individuals from the study population are randomized at baseline $g_1$ or $g_0$, everyone adheres to the protocol prescribed by the randomization and the outcome is completely and accurately measured at $t^*$.  Third,  we emphasize that, along with the choices of $g_1$ and $g_0$ the investigator chooses to compare, the choice of $t^*$ indexing the time at which this comparison will be made is a defining feature of the target causal effect \eqref{total}.  Elucidating the investigator's underlying causal story (probing what they wish to learn from this study and why) can be essential for separating the actual causal question of interest from data that happens to be available in the analytic data set \cite{young2024story}. This is particularly the case for settings with repeatedly measured outcomes. For example, the observed data structure described in Section \ref{sec: obsdata} includes measures of weight change up to 24 months after treatment initiation.  Without engagement with the details of why this study is being conducted and its goals for treatment guidance and decision making, it would be tempting to assume interest is in the effect \eqref{total} for \textsl{all} $t^*=1,\ldots,24$.  However, engagement with these details clarifies this is not the case.  The actual underlying motivation behind the MedWeight study is that early weight gain is a prominent reason that patients cite as a reason for discontinuing antidepressant treatments after initiating. It turn, while any $t^* \leq 6$ months might be a reasonable translation of this story, $t^*>6$ months may not reasonably align.   Generally, the value(s) of $t^*$ that best align with the underlying causal question needs to be considered on a case-by-case basis based on initial probing conversations with subject matter experts\cite{young2024story}.

\section{Identification when deaths do not occur} \label{sec: identificationtotal}

We noted in the previous section that the total effect \eqref{total} is guaranteed identified in an ``ideal'' trial where the treatment strategies of interest are physically randomized, there is perfect adherence and complete outcome meausurement at $t^*$.  In non-ideal studies such as MedWeight, assumptions will be required that are not guaranteed by design.  In this section we consider more general identification results.  

Theorem 31 of Richardson and Robins \cite{richardson2013single} gives an exchangeability condition for identification of $\mbox{E}[Y_{t^*}^{g}]$ for any $g\in\mathcal{G}$.  This condition can be directly evaluated on a Single World Intervention Graph (SWIG) associated with a ``perturbed'' version of $g$ as follows (Lemma 33 and Corollary 34 of \cite{richardson2013single}): For each time $t=0,\ldots,t^*$, define the SWIG associated with the ``perturbed'' strategy $g(-t)$ as equal to the SWIG associated with $g$ if treatment assignment under $g$ does not depend on natural treatment history.  Otherwise, define the SWIG associated with $g(-t)$ as the SWIG associated with $g$ but with all arrows out of $A^g_t$ (the natural treatment value at $t$ under $g$) removed.  Exchangeability can then be evaluated on this SWIG by: 
\begin{itemize}
\item[(A1).] ``Exchangeability for treatment'': the absence of any unblocked backdoor paths connecting $A^{g(-t)}_t$ and $Y_{t^*}^{g(-t)}$ conditional on $(\overline{L}^{g(-t)}_t, \overline{A}^{g(-t)}_{t-1},\overline{A}^{{g(-t)}+}_{t-1},\overline{X}^{{g(-t)}}_{t-1}$), $t=0,\ldots,t^*$. 
\item[(A2).]``Exchangeability for censoring'': the absence of any unblocked backdoor paths connecting $X^{g(-t)}_t$ and $Y_{t^*}^{g(-t)}$ conditional on $(\overline{L}^{g(-t)}_t, \overline{A}^{g(-t)}_{t},\overline{A}^{{g(-t)}+}_{t},\overline{X}^{{g(-t)}}_{t-1}$), $t=0,\ldots,t^*$.
\end{itemize}
See Richardson and Robins \cite{richardson2013single} for full technical details on construction of SWIGs with examples.  For additional examples, see Young et al. \cite{young2014identification} and Wanis et al. \cite{wanis2024grace}. 

Given (A1) and (A2), along with a consistency condition connecting counterfactual variables indexed by $g$ to factual ones \cite{richardson2013single,young2024story}, we can nonparametrically identify $\mbox{E}[Y_{t^*}^{g}]$ via Robins's generalized g-formula, a function of only the observed data \cite{robins1986new, richardson2013single}:
\begin{align}
   \psi_{t^*}(g)= & \sum_{\overline{l}_{t^*}}\sum_{\overline{a}_{t^*}} \Big\{ \E[Y_{t^*}|\overline{L}_{t^*}=\overline{l}_{t^*},\overline{A}_{t^*}=\overline{a}_{t^*},\overline{X}_{t^*}=0] \nonumber\\
   &\quad \times \prod_{s=0}^{t^*} f^{g}(a_s|\overline{l}_s,\overline{a}_{s-1},\overline{X}_{s}=0) f(l_s|\overline{l}_{s-1},\overline{a}_{s-1},\overline{X}_{s}=0) \Big\}, \label{eq: gform}
\end{align}
provided this function is defined \cite{young2014identification}, where $\overline{X}_0\equiv\overline{A}_{-1}\equiv\overline{L}_{-1}\equiv 0$ and the summations in (\ref{eq: gform}) are over the support of $\overline{L}_{t^*}$ and $\overline{A}_{t^*}$, respectively.  For simplicity, we have written $\psi_{t^*}(g)$ in \eqref{eq: gform} explicitly for the case where $\overline{L}_{t^*}$ and $\overline{A}_{t^*}$ are discrete; otherwise, we can replace the summations with integrals.  Here, $\E[Y_{t^*}|\overline{L}_{t^*}=\overline{l}_{t^*},\overline{A}_{t^*}=\overline{a}_{t^*},\overline{X}_{t^*}=0]$ is the mean of $Y_{t^*}$ conditional on remaining uncensored through $t^*$ and $(\overline{L}_{t^*},\overline{A}_{t^*})$ evaluated at $(\overline{l}_{t^*},\overline{a}_{t^*})$. For time $s=0,\ldots, t^*$, $f(l_s|\overline{l}_{s-1},\overline{a}_{s-1},\overline{X}_{s}=0)$ is the density of $L_s$ conditional on remaining uncensored through $s$ and $(\overline{L}_{s-1},\overline{A}_{s-1})$ evaluated at $(\overline{l}_{s-1},\overline{a}_{s-1})$.  Following arguments in Section 9 of Richardson and Robins \cite{richardson2013single}, identification of $\mbox{E}[Y_{t^*}^{g}]$ by the generalized g-formula \eqref{eq: gform} under an agnostic causal model that does not postulate the existence of individual-level counterfactuals can be evaluated in a graph isomorphic to the SWIG referenced in conditions (A1) and (A2). For $s=0,\ldots, t^*$, we refer to $f^{g}(a_s|\overline{l}_s,\overline{a}_{s-1},\overline{X}_{s}=0):=f_{A_s^{g+}|\overline{L^g_s},\overline{A}^{g+}_{s-1},\overline{X}^{g+}_s}(a_s|\overline{l}_s,\overline{a}_{s-1},0)$ in expression \eqref{eq: gform} as the \textsl{intervention treatment density} associated with $g$ at time $s$, which is the density of $A^{g+}_s$ conditional on remaining uncensored through $s$ and $(\overline{L}^g_{s},\overline{A}^{g+}_{s-1})$ evaluated at $(\overline{l}_s, \overline{a}_{s})$.  Importantly, the intervention treatment density indexing the generalized g-formula \eqref{eq: gform} is marginal with respect to the history of natural treatment values under the selected $g\in \mathcal{G}$ even when treatment assignment under $g$ depends on this history.   

A sufficient positivity condition ensuring the function \eqref{eq: gform} is defined for any $g\in \mathcal{G}$ can be stated as follows \cite{young2014identification}
\begin{itemize}
\item[(A3).] Positivity for $g$: For $t=0,\ldots,t^*$
\begin{align*}
&\mbox{if }f(\overline{l}_t,\overline{a}_{t-1},\overline{X}_{t-1}=0)>0\mbox{ then }\\
&f^{g}(a_t \mid \overline{l}_t,\overline{a}_{t-1}, \overline{X}_{t-1}=0) > 0 \implies f^{obs}(a_t \mid \overline{l}_t, \overline{a}_{t-1}, \overline{X}_{t-1}=0) > 0
 \end{align*}
where we refer to $f^{obs}(a_t|\overline{l}_t,\overline{a}_{t-1},\overline{X}_{t}=0):=f_{A_t|\overline{L_t},\overline{A}_{t-1},\overline{X}_t}(a_t|\overline{l}_t,\overline{a}_{t-1},0)$ as the \textsl{observed treatment density}.  In words, this condition requires that, conditional on remaining uncensored and any possibly observed treatment and confounder history, if a treatment level $a_t$ is possible to observe under $g$, it must also be possible to factually observe in the study population. 

\item[(A4).] Positivity for ``censoring elimination'': 
For $t=0,\ldots,t^*$
\begin{equation*}
\mbox{if }f(\overline{l}_t,\overline{a}_{t},\overline{X}_{t-1}=0)>0\mbox{ then }
\Pr(X_t=0 \mid \overline{l}_t, \overline{a}_{t-1}, \overline{X}_{t-1}=0) > 0
 \end{equation*}
 In words, this condition requires that conditional on remaining uncensored and any possibly observed treatment and confounder history, there is a positivity probability of remaining uncensored through that time.     
\end{itemize}

Note that conditions (A1)-(A4) are all dependent on the choice of outcome time $t^*$, with the number of restrictions they require increasing for larger choices of $t^*$.  Thus, in addition to the meaning of the causal effect \eqref{total} and its relevance to the underlying study purpose and motivation, our ability to reasonably identify this effect with only measured variables is fundamentally dependent on the choice of outcome time $t^*$.

\section{An inverse probability weighted representation of the generalized g-formula} \label{sec: ipw representation}
Let $V$ denote an investigator-selected subset of the baseline covariates $L_0$, which can include the empty set or all of $L_0$.  In turn, denote $\psi_{t^*}(g,V)$ as the generalized g-formula $\eqref{eq: gform}$ but conditioned on the chosen subset $V$.  For our subsequent construction of an estimator of the generalized g-formula \eqref{eq: gform}, it will be useful to note that $\psi_{t^*}(g,V)$ has the following alternative, yet equivalent, inverse probability (IP) weighted mean representation: 
\begin{equation} \label{eq: ipw}
   \psi_{t^*}(g,V)=\mbox{E}[Y_{t^*}W^{g}_{t^*}W^{\overline{x}_{t^*}=0}_{t^*}|V],
\end{equation}
where  
\begin{equation}
   W^{g}_{t^*} = \frac{\prod_{s=0}^{t^*} f^{g}(A_s|\overline{L}_s,\overline{A}_{s-1},\overline{X}_{s-1}=0)}{\prod_{s=0}^{t^*} f^{obs}(A_s|\overline{L}_s,\overline{A}_{s-1},\overline{X}_{s-1}=0)} \label{gweight}
   \end{equation}
 and
   \begin{equation}
    W^{\overline{x}_{t^*}=0}_{t^*} = \frac{\prod_{s=0}^{t^*}I(X_{s}=0)}{ \prod_{s=0}^{t^*}\Pr(X_{s}=0|\overline{L}_{s},\overline{A}_{s},\overline{X}_{s-1}=0)}.\label{censweight}
\end{equation} 
See Appendix A for proof. For our definition of censoring events, we can more explicitly write \eqref{censweight} as the product $W^{\overline{x}_{t^*}=0}_{t^*}=W^{r_{t^*}=1}_{t^*}\times W^{\overline{c}_{t{^*}}=0}_{t^*}$ where
\begin{equation}
    W^{r_{t^*}=1}_{t^*} = \frac{I(R_{t^*}=1)}{ \Pr(R_{t^*}=1|\overline{L}_{t^*},\overline{A}_{t^*},\overline{C}_{t^*}=0)}\label{rweight}
\end{equation} 
and
\begin{equation}
    W^{\overline{c}_{t^*}=0}_{t^*} = \frac{\prod_{s=0}^{t^*}I(C_{s}=0)}{ \prod_{s=0}^{t^*}\Pr(C_{s}=0|\overline{L}_{s},\overline{A}_{s},\overline{C}_{s-1}=0)}\label{cweight}
\end{equation} 

Consider our examples of natural grace period strategies indexed by a grace period of length $m$ for citalopram and sertraline defined in Section \ref{sec: question}.  Following arguments in Wanis et al. \cite{wanis2024grace}, the weight \eqref{gweight} for the citalopram strategy ($g_{1}$) would take the specific form
 \begin{equation*}
        W^{g_{a_0}}_{t^*}  = \frac{I(A_0=a_0)}{\Pr(A_0=a_0|L_0)} \times \prod_{s=1}^{t^{*}} \left\{(1 - G^m_s) + \frac{G^m_s \times A_s}{\Pr(A_s = 1 | G^m_s = 1, \overline{L}_{s},\overline{A}_{s-1},\overline{X}_{s-1}=0)}\right\},
    \end{equation*}
for $a_0=1$.  Analogously, the weight \eqref{gweight} for the sertraline strategy ($g_0$) would take this same form but for $a_0=0$.

\section{A time-smoothed IP weighted estimator of the total effect} \label{sec: time smoothed estimation}
Now let $\psi_t(g,V)$ denote the right hand side of \eqref{eq: ipw} but replacing the outcome time of interest $t^*$ with arbitrary $t=1,\ldots,\tau$. Further, assume there exists a real-valued coefficient vector $\theta_0$ such that the following equality is true for all $t=1,\ldots,\tau$ (i.e., all the times at which at least some individuals have outcome measures in the data to be analyzed), the investigator-choice of $V$, and all $g\in\mathcal{G}^{'}\subseteq \mathcal{G}$:  
\begin{equation} \label{eq: pool}
 \psi_{t}(g,V)= \gamma(g,V,t;\theta_0) 
\end{equation}
where $\gamma(\cdot)$ is a linear function of $t$, $V$, and $g$ and is indexed by coefficients $\theta_0$, and $\mathcal{G}^{'}$ contains only the set of strategies of interest (e.g. antidepressant strategies $g_1$ and $g_0$ defined in Section \ref{sec: question}).  Under the model assumption \eqref{eq: pool}, we then have that the generalized g-formula \eqref{eq: gform} indexed by outcome time of interest $t^*$ can be written as
\begin{equation}
  \psi_{t^*}(g)=\sum_v  \gamma(g,v,t^*;\theta_0)\Pr(V=v).
\end{equation}

Now define $\hat{\theta}$ as the solution to the estimating equation
\begin{equation*}
\sum_{i=1}^{n}\sum_{g\in\mathcal{G}^{'}}\sum_{t=0}^\tau U^{g}_{i,t}(\theta,\hat{\alpha},\hat{\beta},\hat{\phi})=0,
\end{equation*} 
where
\begin{align*}
U^{g}_{i,t}(\theta,\hat{\alpha},\hat{\beta},\hat{\phi})=[Y_{i,t}-\gamma(g,V_{i},t;\theta)]W^g_{t,i}(\hat{\alpha})W^{r_{t}=1}_{t}(\hat{\beta})W^{\overline{c}_{t}=0}_{t}(\hat{\phi})q(g,V_{i},t),
\end{align*}
with $q(\cdot)$ a user-chosen function of, at most, $g,V,t$ (which can trivially be chosen as $q=1$), and $\hat{\alpha}$ $\hat{\beta}$, $\hat{\phi}$ are the maximum likelihood estimates of the model coefficients of regression models used to estimate the weights defined in \eqref{gweight}, \eqref{rweight}, and \eqref{cweight}, respectively, but replacing the $t^*$ index with arbitrary $t$.  Given the model \eqref{eq: pool} is correctly specified, along with the models for estimation of the weights, then $E[U^{g}_{t}(\theta,\alpha_0,\beta_0,\phi_0)]=0$ with $(\alpha_0,\beta_0,\phi_0)$ the true values of the nuisance model parameters \cite{tsiatis2006semiparametric}.  The following estimator of the generalized g-formula $\psi_{t{*}}(g)$ is, in turn, consistent and asymptotically normal 
    \begin{equation}
  \hat{\psi}_{t^*}(g)=\frac{1}{n}\sum_{i=1}^n  \gamma(g,V_i,t^*;\hat{\theta}).\label{ipwest}
\end{equation}
Note that the choice of function $q(\cdot)$ can only affect precision of this estimator.  

The following outlines a general algorithm for calculating \eqref{ipwest}.  The input dataset to this algorithm contains rows for each of the $i=1,\ldots,n$ individuals in the sample, each row indexed by $t=0,\ldots,\tau$. An individual contributes $\tau+1$ rows to the dataset if they are not lost to follow-up; otherwise, they contribute $c+1$ rows if they are lost to follow-up at time $t = c$. The values of $L_{0i}$ repeat on all rows $t$ for person $i$. Time-varying observed variables indexed by $t>0$ will appear on row $t$. Variables encoding the history of a variable through $t-1$ may also appear on row $t$. The algorithm broadly proceeds as follows:
\begin{itemize}
  \item Step 1: Data copies. For each $g\in\mathcal{G}^{'}$ (that is, for each treatment strategy of interest), an identical copy of the data set is made and the copies are stacked. For example, if there are two strategies of interest ($g_1$ and $g_0$), two copies of the data set are made and stacked such that the combined data set has twice as many rows.  A new column is added to this combined data set ($g$) indexing to which copy a row belongs (e.g. the first copy is coded $g=1$ on all rows and the second coded $g=0$ on all rows).
  \item Step 2: IP weight calculation. For each person $i$, on each row $t$, and in each copy $g$, weight estimates $W^g_{t,i}(\hat{\alpha})$, $W^{r_{t}=1}_{t,i}(\hat{\beta})$, and $W^{\overline{c}_{t}=0}_{t,i}(\hat{\phi})$ are calculated, along with the function $q(g,V_{i},t)$ based on how that function is chosen and the choice of $V$.  An overall weight is then calculated for each person-time record in this $g$-specific copy: 
  \begin{equation*}
      \hat{W}^{g,q}_{ti}= W^g_{t,i}(\hat{\alpha})\times W^{r_{t}=1}_{t,i}(\hat{\beta})\times W^{\overline{c}_{t}=0}_{t,i}(\hat{\phi}) \times q(g,V_{i},t).
  \end{equation*}
  \item Step 3: Weighted outcome regression.  A weighted linear outcome regression is then fit with dependent variable $Y_{ti}$, independent variables a user-chosen linear function $\gamma(\cdot)$ of $g$, $t$, and $V_i$ and weights $\hat{W}^{g,q}_{ti}$ as calculated in Step 2.  The estimated coefficients of this model are $\hat{\theta}$.
  \item Step 4: Estimate (``predict'') $\psi_{t^*}(g,V_i)$  for each $i=1,\ldots,n$.  
  \begin{itemize}
      \item For $V$ empty, obtain a prediction from the model fit in Step 3 for each $g$ and $t^*$ of interest.  This is the final estimate of $\hat{\psi}_{t^*}(g)$ and Step 5 can be skipped.
      \item For $V$ not empty, remove all rows from each copy in the combined data indexed by $t>1$ (keep only the first row for each copy).  Then, for each person $i$, obtain a prediction from the model fit in Step 3 for each $g$ and $t^*$ of interest. 
  \end{itemize}  
  \item Step 5: Average the predictions.  For $V$ not empty, average the predictions for each $g$ and $t^*$ of interest over the $n$ individuals.  This is the final estimate of $\hat{\psi}_{t^*}(g)$.  
  \item Step 6: Causal effect estimate at $t^*$.  The final causal effect estimate can be obtained by a contrast of $\hat{\psi}_{t^*}(g)$ for any two choices of $g\in \mathcal{G}^{'}$.  For example, an estimate of  \eqref{total} is simply the difference  $\hat{\psi}_{t^*}(g_1)-\hat{\psi}_{t^*}(g_0)$.
\end{itemize}
Confidence intervals can be constructed by bootstrapping the $n$ original IDs $B$ times (e.g. $B=10,000)$, repeating all steps above in each bootstrap sample. For example, a 95\% confidence interval based on the percentile method can be obtained by taking the 2.5 and 97.5 quantiles of the bootstrap distribution \cite{efron1994introduction}.

Note that for certain comparisons (certain choices of $\mathcal{G}^{'}$) the general algorithm above simplifies and the step of making copies is unnecessary. This will be the case when $W^g_{t=0,i}$ cannot be positive for both $g_1$ and $g_0$.  This is the case in our example antidepressant strategies.  An individual who initiates citalopram monotherapy will have a positive value of $W^{g_1}_{t=0,i}$ but then must have $W^{g_0}_{t=0,i}=0$.  In this case, making copies in Step 1 is unnecessary because in Step 2, an individual will receive 0 weight on all rows of all copies other than the one that their data is consistent with at baseline. In Appendix F, we describe more detailed implementations of the above algorithm for these particular strategies. See Cain et al. \cite{cain2010start} for an example where interest is in comparing strategies $g$ where making copies is necessary/the algorithm does not simplify. 

\subsection{An observed data model versus a marginal structural model}\label{pseudomsm}
The model \eqref{eq: pool} is a parametric model for the generalized g-formula itself conditional on a chosen subset $V$ (possibly empty) of the measured baseline covariates indexed by outcome time $t$ and strategy $g$, as a function of $g,t,V$.  That is, the model \eqref{eq: pool}
imposes a restrictive assumption on the joint distribution of the observed data. This model does not necessarily imply the following marginal structural (causal) model assumption:
\begin{equation}
\mbox{E}[Y^{g}_{t}|V]= \gamma(g,V,t;\theta_0),  \label{truemsm}
\end{equation}
for \textsl{all} $t = 1, \dots, \tau$.  Rather, under the additional (causal) nonparametric identifying assumptions (A1)-(A4) of Section \ref{sec: identificationtotal},  the model \eqref{eq: pool}
 implies the equality \eqref{truemsm} \textsl{only} for the selected outcome time(s) of interest $t=t^*$. In the special case where these outcome times of interest happen to align with all times at which outcome measures are available in the data $t^*=1,\ldots,\tau$ \textsl{and} we are willing to rely on the causal assumptions of Section \ref{sec: identificationtotal} for all of these choices of $t^*$ then the model \eqref{eq: pool} can be understood as a marginal structural model.  Otherwise, the model \eqref{eq: pool} is more accurately understood as a ``statistical'' assumption - an assumption imposed on the observed data for the purpose of leveraging all outcome measures for precision gain.  This assumption will generally come with a price of some increased bias.  We illustrate this tradeoff via simulations in Section \eqref{sec: simulations}.

Finally, note that the IPW estimator above can easily be adapted to remove dependence of the model \eqref{eq: pool} on the treatment strategy $g$ and thus avoid any parametric assumptions on the nature of dependence between $g$ and the outcome at any time.  In this case, Step 3 can be modified to fit separate weighted outcome regression models to each g-specific copy where the independent variables in each copy include only a function of $t$ and $V$; that is, the stacked data set can be stratified on the values of the column ``$g$'' created in Step 1. Then in Step 4, estimates  associated with a strategy $g$ would be made from the model in Step 3 specific to that value of column ``$g$'' in the stacked data set.

\section{Defining and interpreting a ``story-led'' notion of causal effect when deaths occur} \label{sec: identification deaths}

We now relax the assumption that no deaths can occur during the study follow-up in this population.  Specifically, suppose that 
$\Pr(D_t^{g}=1)>0$ for any $g\in\mathcal{G}^{'}$, 
$0<t<t^*$ and $t^*$ of interest.    In this case, the total effect \eqref{total} premising all of the arguments above is undefined and a different notion of causal effect must be chosen to ground causal inferences.

In this case, the most commonly considered effect, particularly, in the setting of generalized treatment strategies/LMTPs, coincides with \eqref{total} but where death is considered another source of censoring.  In turn, \eqref{total}, which refers to an effect under ``censoring elimination'', additionally references an effect under ``death elimination".  Such an effect can be understood as a case of a controlled direct effect from the mediation literature where the death process can be viewed as a time-varying mediator \cite{robins1992identifiability, young2020causal}. In this case, some direct or path-specific effect that somehow isolates treatment mechanism is inherently appealing because the total effect cannot be (at least meaningfully) defined.  Identification and estimation of such a controlled direct effect immediately follows from arguments in Sections \ref{sec: identificationtotal}, \ref{sec: ipw representation}, and \ref{sec: time smoothed estimation} by allowing the  censoring indicator history $\overline{X}_t$ to additionally include death $\overline{D}_t$.  Despite the simplicity of this effect choice in terms of easy adaptation of existing results for ``right-censored data'', effects that include reference to interventions that somehow universally eliminate death rarely align with what is actually motivating an investigator, particularly in a study of a clinical population where the goal is to inform real-world treatment decisions.  

\begin{figure}[H]
\centering
\begin{tabular}{cc}

% Panel A
\begin{subfigure}[t]{0.45\textwidth}
\begin{tikzpicture}
\begin{scope}[every node/.style={thick,draw=none}]
    \node (U2) at (5,-2) {$U_{DY}$};
    \node (A0) at (0,0) {$A_0|A_0^{g+}$};
    \node (A) at (2,0) {$A_s^{g}|A_s^{g+}$};
    \node (D) at (4,0) [draw,thick,minimum width=1cm,minimum height=1cm] {$D_{t^*}^{g}$};
    \node (Y) at (6,0) {$Y_{t^*}^{g}$};
\end{scope}

\begin{scope}[>={Stealth[black]},
              every node/.style={fill=white,circle},
              every edge/.style={draw=black,very thick}]
    \path [->] (U2) edge (D);
    \path [->] (U2) edge (Y);
    \path [->] (A0) edge (A);
    \path [->] (A) edge[bend left] (Y);
    \path [->] (A0) edge[bend left] (Y);
    \path [->] (D) edge (Y);
\end{scope}
\end{tikzpicture} 
\caption{No violation of exchangeability; No violation of treatment-death isolation}
\end{subfigure}

& 

% Panel B
\begin{subfigure}[t]{0.45\textwidth}
\begin{tikzpicture}
\begin{scope}[every node/.style={thick,draw=none}]
    \node (U2) at (5,-2) {$U_{DY}$};
    \node (A0) at (0,0) {$A_0|A_0^{g+}$};
    \node (A) at (2,0) {$A_s^{g}|A_s^{g+}$};
    \node (D) at (4,0) [draw,thick,minimum width=1cm,minimum height=1cm] {$D_{t^*}^{g}$};
    \node (Y) at (6,0) {$Y_{t^*}^{g}$};
\end{scope}

\begin{scope}[>={Stealth[black]},
              every node/.style={fill=white,circle},
              every edge/.style={draw=black,very thick}]
    \path [->] (A0) edge[bend left] (D);
    \path [->] (A0) edge (A);
    \path [->] (A) edge (D);
    \path [->] (U2) edge (D);
    \path [->] (U2) edge (Y);
    \path [->] (A) edge[bend left] (Y);
    \path [->] (A0) edge[bend left] (Y);
    \path [->] (D) edge (Y);
\end{scope}
\end{tikzpicture} 
\caption{No violation of exchangeability; Violation of treatment-death isolation}
\end{subfigure} \\
% Panel C
\begin{subfigure}[t]{0.45\textwidth}
\begin{tikzpicture}
\begin{scope}[every node/.style={thick,draw=none}]
    \node (blank) at (-0.45,-0.25) {};
    \node (U1) at (2,-2) {$U_{ZD}$};
    \node (U2) at (5,-2) {$U_{DY}$};
    \node (A0) at (0,0) {$A_0|A_0^{g+}$};
    \node (A) at (2,0) {$A_s^{g}|A_s^{g+}$};
    \node (D) at (4,0) [draw,thick,minimum width=1cm,minimum height=1cm] {$D_{t^*}^{g}$};
    \node (Y) at (6,0) {$Y_{t^*}^{g}$};
\end{scope}

\begin{scope}[>={Stealth[black]},
              every node/.style={fill=white,circle},
              every edge/.style={draw=black,very thick}]
    \path [->] (A0) edge (A);
    \path [->] (U1) edge (blank);
    \path [->] (U1) edge (D);
    \path [->] (U2) edge (D);
    \path [->] (U2) edge (Y);
    \path [->] (A) edge[bend left] (Y);
    \path [->] (A0) edge[bend left] (Y);
    \path [->] (D) edge (Y);
\end{scope}
\end{tikzpicture} 
\caption{Violation of exchangeability; No violation of treatment-death isolation.}
\end{subfigure}

& 

% Panel D
\begin{subfigure}[t]{0.45\textwidth}
\begin{tikzpicture}
\begin{scope}[every node/.style={thick,draw=none}]
    \node (blank) at (-0.45,-0.25) {};
    \node (U1) at (2,-2) {$U_{ZD}$};
    \node (U2) at (5,-2) {$U_{DY}$};
    \node (A0) at (0,0) {$A_0|A_0^{g+}$};
    \node (A) at (2,0) {$A_s^{g}|A_s^{g+}$};
    \node (D) at (4,0) [draw,thick,minimum width=1cm,minimum height=1cm] {$D_{t^*}^{g}$};
    \node (Y) at (6,0) {$Y_{t^*}^{g}$};
\end{scope}

\begin{scope}[>={Stealth[black]},
              every node/.style={fill=white,circle},
              every edge/.style={draw=black,very thick}]
    \path [->] (A0) edge (A);
    \path [->] (A0) edge[bend left] (D);
    \path [->] (A) edge (D);
    \path [->] (U1) edge (blank);
    \path [->] (U1) edge (D);
    \path [->] (U2) edge (D);
    \path [->] (U2) edge (Y);
    \path [->] (A) edge[bend left] (Y);
    \path [->] (A0) edge[bend left] (Y);
    \path [->] (D) edge (Y);
\end{scope}
\end{tikzpicture} 
\caption{Violation of exchangeability; Violation of treatment-death isolation}
\end{subfigure}

\end{tabular}
\caption{Simplified single world intervention graphs (SWIGs) depicting scenarios with and without violations of exchangeability and the treatment-death isolation condition. Variables $U_{ZD}$ and $U_{DY}$ are unmeasured.} \label{fig:swigs}
\end{figure}
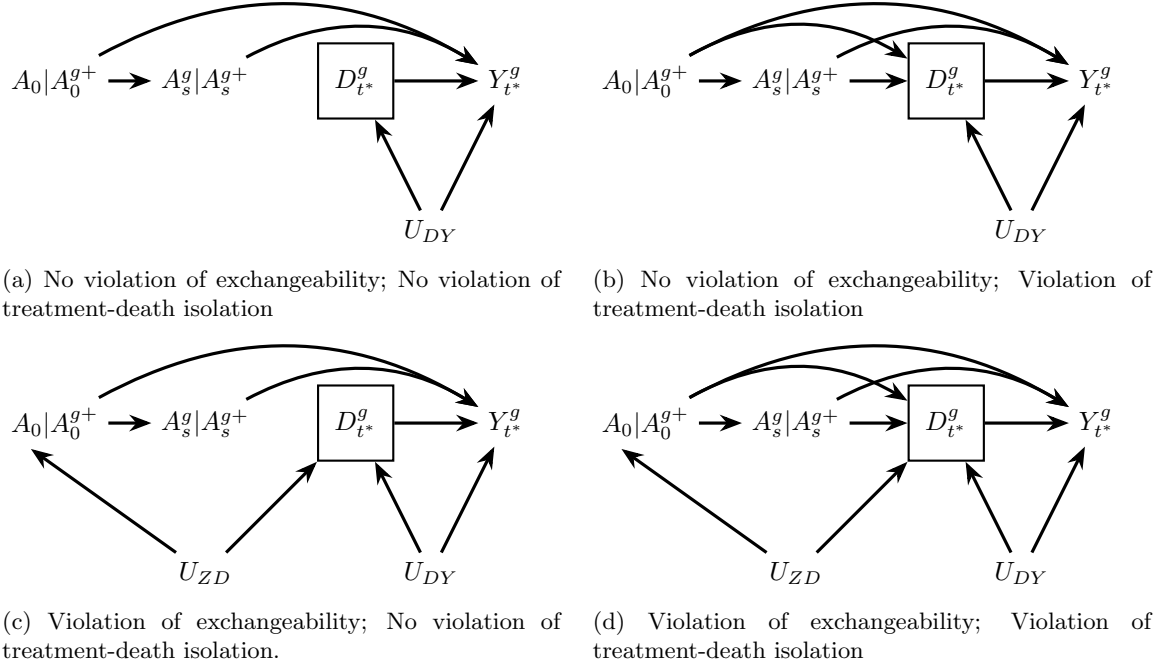

Instead, consider the following alternative contrast in conditional outcome means,  
\begin{equation}\label{eq: estimand deaths}
    \E[Y^{g_1}_{t^{*}} | D^{g_1}_{t^*} = 0] - \E[Y^{g_0}_{t^{*}} | D^{g_0}_{t^*} = 0].
\end{equation}
Like the total effect in our simpler scenario without deaths, and unlike a controlled direct effect under ``death elimination'', the contrast \eqref{eq: estimand deaths} is  guaranteed identified in a randomized trial of the strategies $g_1$ and $g_0$ that is ``ideally'' executed in the sense of perfect adherence, complete capture of survival, and complete outcome measurement among survivors in each treatment arm by $t^*$.  However, unlike the total effect and the controlled direct effect, the effect \eqref{eq: estimand deaths} does not compare outcomes under different treatment strategies in the same population even under a counterfactual ontology: by definition, it compares survivors by $t^*$ under $g_1$ to survivors by $t^*$ under $g_0$.  From the agnostic perspective, \eqref{eq: estimand deaths}, in general, fails to compare outcomes in ``comparable'' populations, even in the ideal trial in which it is guaranteed identified.  This ``non-comparability'' is graphically illustrated in the simplified SWIGs depicted in Figures \ref{fig:swigs}b and \ref{fig:swigs}d by the open (d-connected) ``non-causal'' paths between intervention treatment nodes $A^{g+}_0$ and $A^{g+}_s$ and $Y^g_{t^*}$ conditional on $D^g_{t^*}$, a common effect of the intervention $g$ and an unmeasured shared cause of the outcome $U_{DY}$ and survival and a collider along the paths from $A^{g+}_0$ and $A^{g+}_s$, respectively, to $D^g_{t^*}$ to $U_{DY}$ to $Y_{t^*}$.  The directed arrow from  $D^g_{t^*}$ to $Y^g_{t^*}$ reflects the dependence that if an individual dies their outcome is subsequently undefined\cite{young2021identified, stensrud2023conditional}.  This non-comparability of treatment groups after conditioning on a collider has been referred to as a type of ``selection bias'' \cite{hernan2004structural}.

Consider the following graphical condition relative to a SWIG associated with any $g\in \mathcal{G}$ representing the investigator's causal model:
\begin{itemize}
    \item[(B6)] There are no directed paths connecting $A^{g+}_s$ and $Y^g_{t^*}$ that are intersected by $D^g_t$ for $s<t<t^*$, $s=0,\ldots,t-1, t=1\ldots,t^*$.
\end{itemize}
Inspired by Stensrud et al.\cite{stensrud2023conditional}, we will refer to B6 as the assumption of \textsl{treatment-death isolation}.  The isolation condition B6 can be understood as a trivial case of Stensrud et al.'s full isolation condition on treatment components and, by extension, the contrast \eqref{eq: estimand deaths} a trivial case of a conditional separable effect of generalized time-varying strategies\cite{stensrud2023conditional}.  The isolation condition B6 ensures that the contrast \eqref{eq: estimand deaths} compares outcomes under different treatment strategies across ``comparable'' study populations (i.e. is not subject to the ``selection bias'' induced by conditioning on a collider).  The assumption B6 is an assumption on treatment mechanism.  In addition to allowing interpretation of the contrast \eqref{eq: estimand deaths} as a causal effect in the sense of comparing ``comparable'' populations, a byproduct of B6 is that this effect does capture isolated treatment mechanisms outside of survival. We saw that the definition and identification of the total effect is fundamentally dependent on the choice of $t^*$.  The choice of $t^*$ not only implicates the definition and identification of the estimand \eqref{eq: estimand deaths} but also the viability of B6 impacting interpretation.  We have already established that the ``causal story'' behind the MedWeight study supports interest in choices of $t^*$ early in the study period, closer to baseline (e.g. $t^*\leq 6$ months.  In our antidepressant example, assumption B6 will also be most reasonable in this short initial period post-baseline (i.e for smaller choices of $t^*$) than for larger choices.  

Under a stronger interpretation of the absence of an arrow on SWIG as an assumption of no individual level effect B6 implies $D_{t^*}^{g} = D_{t^*}$ for any strategy $g \in \mathcal{G}^\prime \subseteq \mathcal{G}$ and the contrast \eqref{eq: estimand deaths} will coincide with a case of a survivor average causal effect.  

\section{Identification under treatment-death isolation}

Following argument above, a conditional outcome mean $\E[Y^{g}_{t^{*}} | D^{g}_{t^*} = 0]$ is guaranteed identified in an ``ideal'' trial.  In such a trial, the isolation condition B6 is only required to interpret the contrast \eqref{eq: estimand deaths} as a causal effect; a comparison of outcomes under different treatment strategies across ``comparable'' populations using our graphical understanding of ``comparable''.  

In more general, ``non-ideal'' studies, we can identify $\E[Y^{g_z}_{t^{*}} | D^{g_z}_{t^*} = 0]$ under a modified version of the identifying conditions (A1)-(A4) defined in Section \ref{sec: identificationtotal} to allow additional conditioning on survival through $s$, coupled with additional treatment and censoring exchangeability conditions referencing survival in place of the outcome of interest. See Appendix D.

Under these conditions, it can be shown that we can identify  $\E[Y^{g}_{t^{*}} | D^{g}_{t^*} = 0]$ by
\begin{align}
   & \sum_{\overline{l}_{t^*}}\sum_{\overline{a}_{t^*}} \Big\{ \E[Y_{t^*}|\overline{L}_{t^*}=\overline{l}_{t^*},\overline{A}_{t^*}=\overline{a}_{t^*},\overline{X}_{t^*}=0, \overline{D}_{t^*}=0] \nonumber\\
   &\qquad \times \prod_{s=0}^{t^*} \Pr(D_s = 0|\overline{L}_{s}=\overline{l}_{s},\overline{A}_{s}=\overline{a}_{s},\overline{X}_{s-1}=0,\overline{D}_{s-1}=0) \nonumber\\
   & \qquad \times \prod_{s=0}^{t^*} f^{g}(a_s|\overline{l}_s,\overline{a}_{s-1},\overline{X}_{s-1}=0,\overline{D}_{s-1}=0) f(l_t|\overline{l}_{s-1},\overline{a}_{s-1},\overline{X}_{s-1}=0,\overline{D}_{s-1}=0) \Big\} \nonumber\\ & \times \Bigg[ \sum_{\overline{l}_{t^*}}\sum_{\overline{a}_{t^*}} \prod_{s=0}^{t^*} \Pr(D_s = 0|\overline{L}_{s}=\overline{l}_{s},\overline{A}_{s}=\overline{a}_{s},\overline{X}_{s-1}=0,\overline{D}_{s-1}=0) \nonumber\\
   & \qquad \times \prod_{s=0}^{t^*} f^{g}(a_s|\overline{l}_s,\overline{a}_{s-1},\overline{X}_{s-1}=0,\overline{D}_{s-1}=0)  f(l_s|\overline{l}_{s-1},\overline{a}_{s-1},\overline{X}_{s-1}=0,\overline{D}_{s-1}=0) \Big\} \Bigg]^{-1},\label{eq: gform deaths v1}
\end{align}
where $f(l_s|\overline{l}_{s-1},\overline{a}_{s-1},\overline{X}_{s-1}=0,\overline{D}_{s-1}=0)$ is taken to mean $f(l_0)$ at time $s=0$. With the additional assumption B6, the identification formula in (\ref{eq: gform deaths v1}) simplifies to 
\begin{align}
   \psi_{t^*}(g, D_{t^*} = 0)= & \sum_{\overline{l}_{t^*}}\sum_{\overline{a}_{t^*}} \Big\{ \E[Y_{t^*}|\overline{L}_{t^*}=\overline{l}_{t^*},\overline{A}_{t^*}=\overline{a}_{t^*},X_{t^*}=0, \overline{D}_{t^*}=0] \nonumber\\
   &\quad \times \prod_{s=0}^{t^*} f^{g}(a_s|\overline{l}_s,\overline{a}_{s-1},X_{s-1}=0,\overline{D}_{t^*}=0)  f(l_s|\overline{l}_{s-1},\overline{a}_{s-1},X_{s-1}=0,\overline{D}_{t^*}=0) \Big\} \label{eq: gform deaths},
\end{align}
where $f(l_s|\overline{l}_{s-1},\overline{a}_{s-1},X_{s-1}=0,\overline{D}_{t^*}=0)$ at $s=0$ is taken to mean $f(l_0|D_{t^*}=0)$. Observe that this is the same identification formula as in (\ref{eq: gform}) except that all terms within the formula are conditioned on $D_{t^*} = 0$.

\section{Time-smoothed IPW estimation under treatment-death isolation} \label{sec: methods deaths}

In this section, we present two possible IPW estimators of the observed data function that can leverage time-smoothing in available outcome measurements.

\subsection{Nonstacked, time-smoothed IPW estimator} 

For a choice of $t^*$ define
\begin{align*}
   \psi_{t}(g, D_{t^*} = 0) := & \sum_{\overline{l}_{t}}\sum_{\overline{a}_{t}} \Big\{ \E[Y_{t}|\overline{L}_{t}=\overline{l}_{t},\overline{A}_{t}=\overline{a}_{t},Z=z,X_{t}=0, \overline{D}_{t^*}=0] \\
   &\quad \times \prod_{s=0}^{t} f^{g_z}(a_s|\overline{l}_s,\overline{a}_{s-1},X_{s-1}=0,\overline{D}_{t^*}=0) f(l_s|\overline{l}_{s-1},\overline{a}_{s-1},X_{s-1}=0,\overline{D}_{t^*}=0) \Big\}.
\end{align*}
for each $t=1,\ldots,t^*$.  Note that $ \psi_{t=t^*}(g, D_{t^*} = 0)$ is the identifying function \eqref{eq: gform deaths}.  Following similar logic to to Section \ref{sec: ipw representation}, we can re-express $\psi_{t}(g, D_{t^*} = 0)$ by
\begin{equation*}
   \psi_{t}(g, D_{t^*} = 0)= \sum_{v}\psi_{t}(g,v, D_{t^*} = 0)\Pr(v|D_{t^{*}}=0).
\end{equation*}
where
\begin{equation*}
    \psi_{t}(g,V, D_{t^*} = 0)\equiv \mbox{E}[Y_{t}W^{g}_{t}W^{x_{t}=0}_{t}|V,D_{t^*} = 0]
\end{equation*}
with weights $W^{g}_{t}$ and $W^{\overline{x}_{t}=0}_{t}=W^{r_{t}=1}_{t}\times W^{\overline{c}_{t{}}=0}_{t}$ defined as in Section \ref{sec: ipw representation}. 

In this time-smoothing strategy, we adopt the following model for all $g \in \mathcal{G}^\prime \subseteq \mathcal{G}$, $t=1,\ldots,t^*$
\begin{equation*}
    \psi_{t}(g,V, D_{t^*} = 0)= \gamma^{\mathrm{nonstacked}}(g,V,t;\theta_0)
\end{equation*}
where $\gamma^{\mathrm{nonstacked}}(\cdot)$ is some function of $t$, $V$, and $g$ and is indexed by coefficients $\theta_0$.  The estimator of $\psi_{t^*}(g,D_{t^*} = 0)$ is then given by
\begin{equation*}
    \hat{\psi}^{\mathrm{nonstacked}}_{t^*}(g,D_{t^*} = 0) = \frac{1}{|\mathcal{S}_{t^*}|} \sum_{i \in \mathcal{S}_{t^*}}  \gamma^{\mathrm{nonstacked}}(g,V_i,t^*;\hat{\theta})
\end{equation*}
where $\mathcal{S}_{t^*} = \{i \in \{1, \dots, n\}: D_{t^*,i} = 0\}$. This estimator can be seen to be the same as the time-smoothed IPW estimator in settings without death after removing all records corresponding to $t>t^*$ and removing all individuals with $D_{t^*} = 1$. We outline the estimation algorithm for our running example in Appendix E. This approach has the advantage of computational simplicity but may have limited precision benefits from time-smoothing when $t^*$ is selected to be small relative to $\tau$.  Returning to our MedWeight example, if we select $t^*=6$ all available weight change measurements between month 7 and $\tau=24$ will not be used in the analysis.  In the next section, we pose an alternative approach to time-smoothing in this case which can leverage all available outcome measures.

\subsection{Stacked, time-smoothed IPW estimator}
 For any $t \in \{1, \dots, \tau\}$, let $\psi_{t}(g, D_t = 0)$ define:
\begin{align*}
   \psi_{t}(g, D_{t} = 0) = & \sum_{\overline{l}_{t}}\sum_{\overline{a}_{t}} \Big\{ \E[Y_{t}|\overline{L}_{t}=\overline{l}_{t},\overline{A}_{t}=\overline{a}_{t},X_{t}=0, \overline{D}_{t}=0] \\
   &\quad \times \prod_{s=0}^{t} f^{g_z}(a_s|\overline{l}_s,\overline{a}_{s-1},X_{s-1}=0,\overline{D}_{t}=0) f(l_s|\overline{l}_{s-1},\overline{a}_{s-1},X_{s-1}=0,\overline{D}_{t}=0) \Big\}.
\end{align*}
such that $ \psi_{t=t^*}(g, D_{t=t^*} = 0)$ is the identifying function \eqref{eq: gform deaths}. Similar to the previous developments, let $\psi_{t}(g, V, D_{t} = 0)$ denote the function $\psi_{t}(g, D_{t} = 0)$ conditioned in $V$, which can be seen to equal the weighted outcome mean
\begin{equation*}
    \psi_{t}(g,V, D_{t} = 0)\equiv \mbox{E}[Y_{t}W^{g}_{t}W^{x_{t}=0}_{t}|V,D_{t} = 0].
\end{equation*}

This time smoothing approach considers the following model for all $g \in \mathcal{G}^\prime \subseteq \mathcal{G}$, $t=1,\ldots,\tau$:
\begin{equation*}
    \psi_{t}(g,V, D_t = 0)= \gamma^{\mathrm{stacked}}(g,V,t;\theta_0)
\end{equation*} 
for some function $\gamma^{\mathrm{stacked}}(\cdot)$ of $t$, $V$, and $g$ with parameters $\theta_0$. 

Because the model $\gamma^{\mathrm{stacked}}(\cdot)$ conditions on different subsets of individuals ($D_t = 0$) at each time point $t$, the estimation algorithm fits the outcome model on a data set formed by \emph{stacking} time $t$-specific data sets. This motivates the term \emph{stacked, time-smoothed IPW estimator} for this approach.

We next outline a general estimation algorithm for this method. The input dataset format is structured in the same manner as in the case without deaths. The only difference is that, if an individual dies at time $t = d$, the dataset will not contain any rows for that individual at subsequent times $t >d$. The algorithm consists of the following steps:
\begin{itemize}

    \item Step 1: Data copies. Follow Step 1 of the algorithm in Section \ref{sec: time smoothed estimation}: For each $g\in\mathcal{G}^{'}$ (that is, for each treatment strategy of interest), an identical copy of the data set is made and the copies are stacked. For example, if there are two strategies of interest ($g_1$ and $g_0$), two copies of the data set are made and stacked such that the combined data set has twice as many rows.  A new column is added to this combined data set ($g$) indexing to which copy a row belongs (e.g. the first copy is coded $g=1$ on all rows and the second coded $g=0$ on all rows). 

    \item Step 2: IP weight calculation. For $t^\prime = 1, \dots, \tau$, perform the following steps:

    \begin{enumerate}
        \item Step 2a: Form a dataset by removing all records with $t > t^\prime$ and all individuals who died by the end of interval $t^\prime$ from the combined data set in Step 1. Denote the resulting data set $\mathcal{D}_{t^\prime}$.

        \item Step 2b: Using dataset $\mathcal{D}_{t^\prime}$, calculate IP weights as in Step 2 in the algorithm  in Section \ref{sec: time smoothed estimation}: For each person $i$, on each row $t$, and in each copy $g$, weight estimates $W^g_{t,i}(\hat{\alpha})$, $W^{r_{t}=1}_{t,i}(\hat{\beta})$, and $W^{\overline{c}_{t}=0}_{t,i}(\hat{\phi})$ are calculated, along with the function $q(g,V_{i},t)$ based on how that function is chosen and the choice of $V$.  An overall weight is then calculated for each person-time record in this $g$-specific copy: 
  \begin{equation*}
      \hat{W}^{g,q}_{ti}= W^g_{t,i}(\hat{\alpha})\times W^{r_{t}=1}_{t,i}(\hat{\beta})\times W^{\overline{c}_{t}=0}_{t,i}(\hat{\phi}) \times q(g,V_{i},t).
  \end{equation*}

    \end{enumerate}
        
    \item Step 3: Create the stacked dataset: Form a dataset by stacking the row corresponding to time $t$ in datasets $\mathcal{D}_t$ for $t = 1, \dots, \tau$. 
    
    \item Steps 4: Weighted outcome regression: Using the stacked dataset from Step 3, fit the weighted outcome regression model with dependent variable $Y_{ti}$, independent variables a user-chosen function $\gamma(\cdot)^{\mathrm{stacked}}$ of $g$, $t$, and $V_i$ and weights $W^{g,q}_{ti}$ as calculated in Step 2.  The estimated coefficients of this model are $\hat{\theta}$.
       
    \item Step 5: Predict the outcome under $g$ at $t^*$ for each $V_i$, $i=1,\ldots,n$:   
  \begin{itemize}
      \item For $V$ empty, obtain a prediction from the model fit in Step 4 for each $g$ and $t^*$ of interest.  This is the final estimate of $\hat{\psi}_{t^*}(g)$ and Step 6 can be skipped.
      \item For $V$ not empty, remove all rows from each copy in the combined data indexed by $t>1$ (keep only the first row for each copy).  Then, for each person with $D_{t^*} = 0$, obtain a prediction from the model fit in Step 4 for each $g$ and $t^*$ of interest. 
  \end{itemize}  
  
   \item Step 6: Average the predictions.  For $V$ not empty, average the predictions for each $g$ and $t^*$ of interest over all individuals with $D_{t^*} = 0$.  This is the final estimate of $\hat{\psi}_{t^*}(g)$.  
  \item Step 7: Causal effect estimate.  The final estimate of \eqref{eq: estimand deaths} is then simply the difference $\hat{\psi}_{t^*}(g_1)-\hat{\psi}_{t^*}(g_0)$
\end{itemize}

\section{Simulations} \label{sec: simulations}

We perform a series of simulations to illustrate properties of the time-smoothed IPW approaches in settings with the aforementioned methodological challenges. In all simulations, we consider the problem of estimating counterfactual means and mean differences of a continuous outcome under time-varying interventions with grace periods in settings with treatment-confounder feedback and substantial outcome missingness. Motivated by the MedWeight study, our simulations adopt the setting of informing comparative effects between medication regimes involving initiating and following a time-varying adherence protocol to the medication initiated. 

In Appendix F, we provide simulations in settings without deaths. We first consider a relatively simple setting where the models for the IPW approach are correctly specified. Consequently, these simulations illustrate how the time-smoothed IPW approach is unbiased while achieving greater efficiency compared to an IPW approach that does not smooth over time, provided that the model assumptions hold. Then, we consider a more realistically complex scenario where there is model misspecification in the time-smoothed IPW approach. In doing so, we illustrate the bias-variance tradeoff between the time-smoothed and non-smoothed approaches. In the this section, we present the simulations in a setting with deaths. 

\subsection{Data generation and estimands}

We simulated 1000 longitudinal data sets, each with $1000$ individuals and $\tau = 24$ follow-up time points (months). Details on the data generating mechanism are provided in Appendix F. In brief, our simulations consider two different medications available. The probability of adherence to medication initiated at baseline ranged from 0.56 to 0.82 at each time point. There was a binary time-varying confounder affected by past treatment (i.e., treatment-confounder feedback). The outcome measurement was missing with probability ranging from 0.56 to 0.82 at each time point. The probability of death by the end of follow-up was approximately 0.70.

For each medication $z \in \{0, 1\}$, we consider the following treatment regime $g_z$: (i) initiate medication $z$ at baseline, (ii) adhere to taking medication $z$ with a grace period of $m=2$ months, and (iii) ensure an outcome measurement at time $t^*$. Our outcome times of interest were set to be month $t^* \in \{6, 12, 18, 24\}$. At each of these times, we estimated the counterfactual outcome means $\E(Y^{g_0}_{t^*} | D_{t^*}^{g_0} = 0)$ and $\E(Y^{g_1}_{t^*} | D_{t^*}^{g_1} = 0)$ as well as the average treatment effect $\E(Y^{g_0}_{t^*} | D_{t^*}^{g_0} = 0) - \E(Y^{g_1}_{t^*} | D_{t^*}^{g_1} = 0)$. The true values of the counterfactual outcome means were computed by Monte Carlo integration with $10,000$ samples, and the true value of the average treatment effect was 0 at each $t^*$.

\subsection{Methods}

We applied the stacked and non-stacked time-smoothed IPW methods. We compared these methods to an analogous approach that does not smooth over time, i.e. only uses outcome measurements at time $t^*$ to estimate $\psi_{t^*}(g)$. Formally, this approach is a special case of the time-smoothed IPW methods when the outcome model is saturated with respect to $t$ (i.e., the model places no restrictions the outcome mean across levels of $t$). We refer to this approach as the \emph{non-smoothed} IPW method.

We provide a detailed description of the estimation algorithms in this setting in Appendix D. We let $V = L_0$ in these methods. For each method, we fit a correctly specified pooled over time logistic regression model for $A_t$ conditional at being at the end of the grace period, 
\begin{equation*}
        \logit(\Pr(A_s = 1 | G_s = 1, \overline{L}_{s} = \overline{\ell}_{s}, \overline{A}_{s-1} = \overline{a}_{s-1}, D_t = 0)) = \alpha_{0} + \alpha_{1} a_0 + \alpha_{2} \ell_s. 
    \end{equation*}
We fit the following correctly specified pooled over time logistic regression model for $R_t$
    \begin{align*}
        \logit(\Pr(R_s = 1 | \overline{L}_{s} = \overline{\ell}_{s}, \overline{A}_{s} = \overline{a}_{s}, D_t = 0)) & = \beta_{0} + \beta_{1} a_{s} + \beta_{2} a_0 + \beta_{3} \ell_{s} 
    \end{align*}
For stabilizing the inverse probability weights (i.e., for obtaining a $q(\cdot)$ function), we fit the following pooled over time logistic regression model:
\begin{equation*}
    \logit(\Pr(R_s = 1 | g_z, L_{0} = \ell_{0}, D_t = 0)) = \beta_{0} + \beta_{1} z + \beta_{2} \ell_{0}. 
\end{equation*}
For the time-smoothed methods, we fit the following (weighted) pooled over time linear regression model for the outcome
\begin{equation*}
        \gamma^{\mathrm{nonstacked}}(g_z, l_0, t; \theta) =\gamma^{\mathrm{stacked}}(g_z, l_0, t; \theta) = \theta_{0} + \theta_{1} z + \theta_{2} l_0 + \theta_{3} t + \theta_4 l_0 z + \theta_5 l_0 t.
\end{equation*}
For the non-smoothed method, we fit a (weighted) saturated linear regression model for the outcome at time $t=t^*$, i.e., 
\begin{equation*}
        \gamma^{\mathrm{nonsmoothed}}(g_z, l_0; \theta) = \theta_{0} + \theta_{1} z + \theta_{2} l_0 + \theta_3 l_0 z.
\end{equation*}
For all three methods, we constructed 95\% confidence intervals around the estimates of the counterfactual outcome means and average treatment effects using nonparametric bootstrap with the percentile method with 250 replicates.

\subsection{Results}

\begin{figure}[H]
  \centering
   \includegraphics[width=0.8\textwidth]{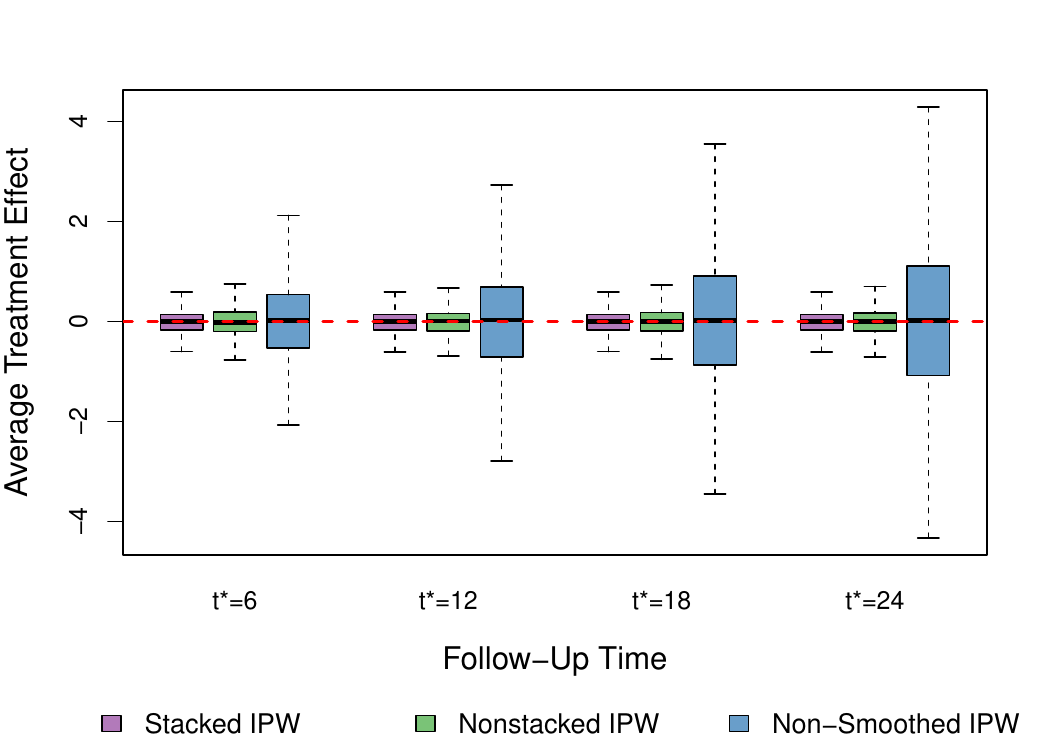}
   \caption{Estimates of the average treatment effect at time point $t^*$, $t^* \in \{6, 12, 18, 24\}$. The dashed red line indicates the true average treatment effect.}\label{fig: sim res deaths}
\end{figure}

Figure \ref{fig: sim res deaths} illustrates the estimates of the average treatment effect at each $t^*$ based on the time-smoothed and non-smoothed IPW estimators. The corresponding bias, standard error, and coverage probabilities are given in Table \ref{tab: sim res deaths}. All three estimators were approximately unbiased and had near nominal coverage of their 95\% CIs. The time-smoothed IPW estimators were considerably more efficient at each time point, with the efficiency gains becoming more pronounced at later time points, e.g. a relative efficiency of approximately 33 for the non-stacked method and 50 for the stacked method at $t^*=24$. The stacked method was more efficient than the non-stacked method at each time point, which may be expected since it involving fitting the outcome model over more data.

The simulation results for the counterfactual outcome means are given in Appendix F. Similar conclusions held for those target parameters: the time-smoothed methods had clear efficiency gains over the non-smoothed method. However, the main difference was that the stacked IPW method underestimated the counterfactual means at earlier time points due to misspecifying the outcome model, which resulted in below nominal coverage at times $t^* = 6$ and $t^* = 12$. The non-smoothed and non-stacked IPW methods were unbiased for the counterfactual means and had close to nominal coverage of their 95\% CIs.

\begin{table}[H]
\caption{Simulation results for estimating the average treatment effect. The bias, standard error, and coverage probability of the 95\% confidence intervals are displayed for time point $t^*$, $t^* \in \{6, 12, 18, 24\}$. The bias and standard error values were multiplied by a factor of 10.}\label{tab: sim res deaths}
\begin{center}
\begin{tabular}{@{\extracolsep{6pt}}llllllllll@{}}
\hline
& \multicolumn{3}{c}{Nonstacked IPW} & \multicolumn{3}{c}{Stacked IPW} & \multicolumn{3}{c}{Non-Smoothed IPW} \\ \cline{2-4} \cline{5-7} \cline{8-10} 
$t^*$ & Bias & SE & Coverage & Bias & SE & Coverage & Bias & SE & Coverage \\  \hline
6 & -0.04 & 2.94 & 0.95 & -0.07 & 2.31 & 0.94 & 0.03 & 7.90 & 0.94 \\ 
  12 & -0.01 & 2.77 & 0.94 & -0.07 & 2.31 & 0.94 & 0.02 & 10.64 & 0.92 \\ 
  18 & -0.07 & 2.77 & 0.95 & -0.06 & 2.31 & 0.94 & 0.59 & 13.08 & 0.94 \\ 
  24 & -0.09 & 2.87 & 0.93 & -0.06 & 2.32 & 0.94 & 0.01 & 16.45 & 0.92 \\ 
   \hline
\end{tabular}
\end{center}
\end{table}

\section{Data illustration} \label{sec: application}

We illustrate applications of the non-smoothed and time-smoothed IPW methods in a re-analysis of the target trial emulation study of Petimar et al.\ \cite{petimar2024medication}. Our analyses estimate comparative effects of medication strategies involving sertraline ($z = 0$) versus citalopram ($z=1$) -- the two most commonly prescribed antidepressants in the dataset -- on weight change. In line with our running example, we consider dynamic, stochastic treatment strategies with the following three components: (i) initiate medication $z$ at baseline, (ii) adhere to taking medication $z$ with a grace period of 2 months, and (iii) ensure a measurement of the outcome at month $t^*$. To illustrate and compare the IPW methods different settings, we consider outcome times $t^*$ of 6, 12, 18, and 24 months and we consider two different outcome variables of interest: weight change from baseline (continuous) and gain 5\% of baseline weight (binary).

\subsection{Dataset and descriptive analyses}

A detailed description of the construction of the dataset we will use is given in Petimar et al.\ \cite{petimar2024medication}. We restricted our analyses to the subgroup of individuals that had a diagnosis of depression and a body mass index (BMI) below 30 kg/m\textsuperscript{2} (i.e., the cutoff for obesity) at baseline. There were several health conditions and medications that may be confounders yet had low prevalence in the dataset (i.e., all below 3\%). To help avoid potential positivity violations, we restricted our analyses to individuals that did not have these health conditions and medications at baseline, which are detailed in Appendix H. 

There were a total of 14,882 individuals in our analytic dataset, of which 7,597 (51.0\%) initiated sertraline and 7,287 (49.0\%) initiated citalopram. The average age was 47 years, 64.1\% were female. Around 79.1\% of individuals where white and 11.4\% were Black/African American. At baseline, 47\% of individuals were overweight (BMI between 25 and 30 kg/m\textsuperscript{2}). Appendix H describes baseline characteristics of individuals that initiated sertraline versus citalopram. 

Outcome measurements were often missing, with 79.1\% missing at 6 months, 80.2\% missing at 12 months, 83.4\% missing at 18 months, and 84.6\% missing at 24 months. Deaths were rare in the dataset: there were 77 deaths (0.5\%) by 6 months, 144 deaths (1.0\%) by 12 months, 191 deaths (1.3\%) by 18 months, and 254 deaths (1.7\%) by 24 months. Overall, 59.0\% of individuals were artificially censored by 6 months, 78.3\% by 12 months, 91.7\% by 18 months, and 94.9\% by 24 months.

\subsection{Methods}

We applied the non-stacked and stacked time-smoothed IPW methods as well as the non-smoothed method (see Section \ref{sec: simulations}) to estimate the counterfactual outcome means (for weight change outcome) and probabilities (for gain of 5\% of baseline weight outcome) $\hat{\E}(Y^{g_z}_{t^*} | D_{t^*}^{g_z} = 0)$, $z \in \{0, 1\}$, $t^* \in \{6, 12, 18, 24\}$. For the weight change outcome, we estimate the counterfactual mean difference $\hat{\E}(Y^{g_1}_{t^*} | D_{t^*}^{g_1} = 0) - \hat{\E}(Y^{g_0}_{t^*} | D_{t^*}^{g_0} = 0)$. For the gain of 5\% of baseline weight outcome, we estimated the counterfactual risk ratio $\hat{\E}(Y^{g_1}_{t^*} | D_{t^*}^{g_1} = 0) / \hat{\E}(Y^{g_0}_{t^*} | D_{t^*}^{g_0} = 0)$

A list of the baseline and time-varying covariates included our analyses is given in Appendix H. These covariates included demographics (e.g., age, sex, race, ethnicity), medical conditions (e.g., anxiety, mental health disorders, type 2 diabetes, hypothyroidism), medications (e.g., antiseizure, antipsychotic, antidiabetic, and antihypertensive prescriptions), lifestyle variables (e.g., smoking status, body mass index), and other variables pertaining to medical history (e.g., Charlson Comorbidity Index). 

In each analysis, we fit a pooled over time logistic regression model for $\Pr(A_s = 1 | G_s = 1, \overline{L}_{s} = \overline{\ell}_{s}, \overline{A}_{s-1} = \overline{a}_{s-1}, D_{t} = 0)$, which included terms for the baseline covariates, time-varying covariates, medication initiated (sertraline vs citalopram), and time (linear and quadratic terms). We fit a pooled over time logistic regression model for the $ \Pr(R_s = 1 | \overline{L}_{s} = \overline{\ell}_{s}, \overline{A}_{s} = \overline{a}_{s}, D_{t} = 0)$, which included the same terms as the treatment model. For obtaining a weight stabilizing $q(\cdot)$ function, we fit a pooled over time logistic regression model for $\Pr(R_s=1|g_z, L_0=l_0,D_{t} = 0)$, which included the baseline covariates, medication initiated, and time. We fit a pooled over time linear (for weight change) or logistic (for a gain of 5\% of baseline weight) regression model for the outcome, which included the baseline covariates, an indicator for antidepressant, and time. Further details on the models are given in Appendix H. We truncated the estimated weights at the 0.99 percentile. We constructed 95\% confidence intervals around our estimates by applying nonparametric bootstrap with the percentile method.

\subsection{Results and interpretation}

The estimates of the counterfactual mean difference in weight change are given in Table \ref{tab: application results continuous v1} and the counterfactual means are given in Appendix H. All methods produced similar point estimates of counterfactual mean difference at each outcome time $t^*$. In particular, all methods estimated a mean difference close to 0 and all confidence intervals included 0. As one would expect, the confidence intervals were considerably more narrow for the time-smoothed methods, especially for the later outcome times. For example, the width of the 95\% confidence interval for the non-smoothed method at $t^* = 24$ was 4.66, whereas the non-stacked and stacked time-smoothed methods had 95\% confidence interval widths of 0.89 and 0.50, respectively.

\begin{table}[H]
\caption{Estimates and 95\% confidence intervals for the counterfactual outcome mean differences in weight change (kg).}\label{tab: application results continuous v1}
\begin{center}
\begin{tabular}{llll}
\hline
& \multicolumn{3}{c}{Counterfactual Mean Difference Estimate (95\% CI)} \\ \cline{2-4}
$t^*$ & Non-Smoothed IPW & Non-Stacked IPW & Stacked IPW \\  \hline
6 & 0.24 (-0.19, 0.70) & 0.09 (-0.08, 0.22) & -0.02 (-0.31, 0.19) \\ 
12 & -0.17 (-1.36, 1.04) & 0.08 (-0.17, 0.27) & -0.02 (-0.31, 0.19) \\ 
18 & -0.80 (-2.28, 1.86) & -0.08 (-0.44, 0.20) & -0.02 (-0.31, 0.19)\\ 
24 & 1.35 (-1.74, 2.92) & -0.16 (-0.64, 0.25) & -0.02 (-0.31, 0.19)\\ 
   \hline
\end{tabular}
\end{center}
\end{table}

Table \ref{tab: application results binary v1} gives the estimates of the counterfactual risk ratios for gaining 5\% of baseline weight, and Appendix H gives the estimates of the counterfactual risks. Similar trends held in these analyses compared to those in the weight change analyses. That is, all methods produced similar counterfactual risk ratio estimates, which were close to 1 at each outcome time $t^*$. The confidence intervals were more narrow for the time-smoothed methods, especially at later outcome times. For example, the width of the 95\% confidence interval for the non-smoothed method at $t^* = 24$ was 0.53, whereas the non-stacked and stacked time-smoothed methods had 95\% confidence interval widths of 0.23 and 0.24, respectively.

\begin{table}[H]
\caption{Estimates and 95\% confidence intervals for the counterfactual risk ratio for gaining 5\% of baseline weight.}\label{tab: application results binary v1}
\begin{center}
\begin{tabular}{llll}
\hline
& \multicolumn{3}{c}{Counterfactual Risk Ratio Estimate (95\% CI)} \\ \cline{2-4}
$t^*$ & Non-Smoothed IPW & Non-Stacked IPW & Stacked IPW \\  \hline
6 & 1.04 (0.89, 1.21) & 1.06 (0.95, 1.22) & 1.06 (0.93, 1.20) \\ 
12 & 1.00 (0.88, 1.24) & 1.07 (0.97, 0.17) & 1.06 (0.95, 1.14) \\ 
18 & 0.74 (0.60, 1.32) & 1.03 (0.94, 1.14) & 1.04 (0.95, 1.13)\\ 
24 & 1.12 (0.83, 1.36) & 1.00 (0.89, 1.12) & 1.05 (0.94, 1.18)\\ 
   \hline
\end{tabular}
\end{center}
\end{table}

Our study population, treatment strategies, target estimands, and statistical methods differ with those in the original analyses by Petimar et al.\ \cite{petimar2024medication}. A key difference in our methodology is how deaths are handled. Nevertheless, our analyses arrived at similar conclusions: weight change trajectories (among survivors) were similar between the medication strategy involving sertraline and that involving citalopram. For each of the strategies, a substantial portion of individuals gained 5\% of baseline weight at each outcome time of interest.

Our re-analysis has several limitations. First, as noted earlier, the causal assumptions required for these methods become stronger as the outcome time $t^*$ increases. We included outcome times ranging from 6 months to 24 months to compare the different IPW methods in various settings; However, the estimates at longer outcome times -- particularly at 24 months -- are especially susceptible to bias. Second, it is likely that some deaths were not captured in the dataset, as death information in EHRs can be incomplete and inconsistently recorded. Third, there is likely some degree of model misspecification in our models for the nuisance parameters, especially in our analyses with longer outcome times. For ease of exposition, our models had the same functional form regardless of the outcome time. In practice, when the outcome time $t^*$, data analysts may wish to use more flexible functions of time in these models, such as including restricted cubic splines for time and interactions with time and medication $z$. 

\section{Discussion} \label{sec: discussion}

Researchers are often interested in estimating causal effects of generalized time-varying treatment strategies. Although measurements of the outcome at intermediate follow-up times are often available, there is limited work in the causal inference literature on methods for leveraging such intermediate outcome data. We developed inverse probability weighted estimators to estimate effects of generalized time-varying treatment strategies that leverage intermediate outcome measurements to improve efficiency. We also consider settings where outcome measurements are truncated by death, which poses challenges in defining meaningful notations of causal effects and developing corresponding estimators. By adopting a treatment-death isolation assumption, we define causal estimands and propose methods that leverage repeated outcome measurements under different types of assumptions on the outcome model. These methods are implemented in the R package \verb|smoothedIPW|, available on GitHub at \url{https://github.com/stmcg/smoothedIPW}.

Our proposal does not require outcome measurements at all time points prior to the follow-up time interval of interest $t^*$, which has some important advantages and disadvantages. As discussed previously, Cole et al.\ \cite{cole2007determining, cole2008constructing} applied inverse probability weighted estimators to estimate effects of treatment strategies that involve eliminating missing outcome measurements at each time point (i.e., eliminating censoring). Such methods would typically face positivity violations in EHR settings such as ours due to the lack of individuals with outcome measurements at each follow-up time interval. However, it is important that not requiring outcome measurements at all time points is subject to unmeasured confounding if past outcome measurements are common causes of treatment and the outcome at the follow-up time of interest. For instance, in our motivating example, an individual becoming aware of their weight gain may affect their continuation of taking their antidepressant as well as their future weight. In practice, data analysts may consider a using a coarse measurement of past outcome values (e.g., weight change in the last 6 time intervals) to mitigate such unmeasured confounding.

Although we considered estimation strategies based on inverse probability weighting, other estimation strategies could be explored in this context. For example, one can construct a parametric g-formula estimator of $\E[Y^{g_z}_{t^{*}} | D^{g_z}_{t^*} = 0]$ based on a plugging in parametric estimates of the conditional means and densities in (\ref{eq: gform deaths}), where one could leverage the repeated outcome measurements by fitting a time-smoothed conditional outcome mean model. We describe such an approach in detail in Appendix J. As discussed previously, a limitation of such an approach is its reliance on a large number of modelling assumptions. Alternatively, one could explore the use of TMLE and SDR estimation methods in this context. We leave such developments for future research endeavors.

\section*{Acknowledgements}

The computations in this paper were run on the FASRC Cannon cluster supported by the FAS Division of Science Research Computing Group at Harvard University.

\bibliographystyle{unsrt}
\bibliography{ref}

\end{document}